\documentclass[aps,pra,floatfix,showpacs,twocolumn,superscriptaddress]{revtex4}

\usepackage{graphics}
\usepackage{graphicx}
\usepackage{epsfig}
\usepackage{amssymb}
\usepackage{amsmath}
\usepackage{amstext}
\usepackage{bm}
\usepackage{times}

\newcommand{\ket}[1]{\ensuremath{|#1\rangle}}

\newcommand{\expect}[1]{\ensuremath{\langle #1 \rangle}}
\newcommand{\ER}{\ensuremath{E_R}}
\newcommand{\intvecr}{\hspace{-0.05cm} \int \hspace{-0.05cm} d^3{\hspace{-0.027cm}}r \hspace{0.1cm} }
\newcommand{\intx}{\int \hspace{-0.5ex} dx\, }
\newcommand{\s}{\ensuremath{\text{s}}}
\newcommand{\as}{\ensuremath{\text{as}}}

\newcommand{\AFFITP}{\affiliation{I. Institut f\"ur Theoretische Physik, Universit\"at Hamburg,
  Jungiusstrasse 9, 20355 Hamburg, Germany} }
\newcommand{\AFFILP}{\affiliation{Institut f\"ur Laser-Physik, Universit\"at Hamburg,
Luruper Chaussee 149, 22761 Hamburg, Germany} }
\newcommand{\AFFMCUA}{\affiliation{Midlands Centre for Ultracold Atoms, School of Physics and Astronomy, University of Birmingham, Edgbaston, Birmingham B15 2TT, United Kingdom} }

\begin{document}
\setlength{\textheight}{24cm}

\title{Localization and delocalization of ultracold bosonic atoms in finite optical lattices}

\author{Dirk-S\"oren L\"uhmann}
\AFFITP
\author{Kai Bongs}
\AFFILP\AFFMCUA
\author{Klaus Sengstock}
\AFFILP
\author{Daniela Pfannkuche}
\AFFITP

\begin{abstract}
We study bosonic atoms in small optical lattices by exact diagonalization and 
observe a striking similarity  to the superfluid to Mott insulator transition in macroscopic systems. 
The momentum distribution, the formation of an energy gap, and the
pair correlation function show only a weak size dependence.
For noncommensurate filling we reveal  in deep lattices   
a mixture of localized and delocalized particles,
which is sensitive to lattice imperfections. 
Breaking the lattice symmetry causes a Bose-glass-like behavior. 
We discuss the nature of excited states and orbital effects 
by using  an exact diagonalization technique that includes higher bands. 
\end{abstract}

\date{19 February 2008}


\maketitle

%%%%%%%%%%%%%%%%%%%%%%%%%%%%%%%%%%%%%%%%%%%%%%%%%%%%%%%%%%%%%%%%%%%%%%%%%%%%%%%%%%%%%%%%%%%%%%%%
%%%%%%%%%%%%%%%%%%%%%%%%%%%%%%%%%%%%%%%%%%%%%%%%%%%%%%%%%%%%%%%%%%%%%%%%%%%%%%%%%%%%%%%%%%%%%%%%

\section{Introduction}

Trapped atoms in optical lattices offer a fantastic new system for applications in 
quantum optics, quantum information processing, and as 
a model system for solid state physics.
The depth and the shape of the lattice potential, 
 which is proportional to the square of the laser field,   
can be controlled with great accuracy \cite{JessenReview}.
Feshbach resonances allow the tuning of the interaction between the neutral atoms in a wide range \cite{Inouye}.
As a fascinating new development, experiments with a small number of lattice sites and
in particular finite optical chains have become focus of actual research. 
Prominent examples are quantum registers in the context of quantum information processing 
\cite{JakschGate,BrennenGate,Schrader},
the manipulation of single atoms within few sites \cite{Miroshnychenko}, 
and experiments with double well unit cells \cite{Sebby}.
In this context a fundamental question arises: 
How similar are finite systems compared with macroscopic systems?

In a macroscopic lattice, bosonic atoms undergo the quantum phase transition 
from a superfluid phase to a Mott insulator 
when the depth of the lattice potential is tuned from shallow to deep. 
This was first discussed for liquid helium on porous media \cite{Fischer}, 
later proposed for neutral repulsively interacting atoms in optical lattices \cite{Jaksch}, and 
recently observed experimentally \cite{Greiner,stoferle:130403,Gerbier}.
Driven by the competition of repulsive interaction and kinetic energy 
the atoms localize on single lattice sites in the Mott insulator  phase. 
Consequently, each site becomes occupied by a fixed integer number of particles leading to a crystal-like situation.
In the presence of disorder the system can also undergo a transition to a Bose glass phase 
\cite{Fischer,scalettar:3144,Damski,Fallani,Pugatch}.

In this article, we study the problem of few repulsively interacting bosonic atoms in finite  linear  chains 
and small two-dimensional lattices and
discuss how finite size effects influence the precursors of the Mott insulator and the Bose glass transition.
The rich physics of the crossover from a double well to mesoscopic systems is investigated by exact 
diagonalization using a multiband basis, which allows accurate results and
the discussion of orbital effects which were widely neglected so far. 

We find a surprisingly strong similarity to the localization process in macroscopic systems
and show that the  pair correlation function  is nearly size independent.
Moreover, we gain an intuitive insight into the excitation spectrum.
Finite systems offer also an unique possibility to study the effects of noncommensurate filling,
in which the localization is suppressed by  the equivalence of the lattice sites.  
We observe the formation of an insulating lowest band  
and  in deep lattices we find  the coexistence of localized and delocalized atoms.
However,  these delocalized states are  extremely sensitive to lattice imperfections 
which force the localization of atoms in a Bose-glass-like phase.

%%%%%%%%%%%%%%%%%%%%%%%%%%%%%%%%%%%%%%%%%%%%%%%%%%%%%%%%%%%%%%%%%%%%%%%%%%%%%%%%%%%%%%%%%%%%%%%%

\section{Theoretical model}

\begin{figure}[t] 
\includegraphics[width=0.9\linewidth]{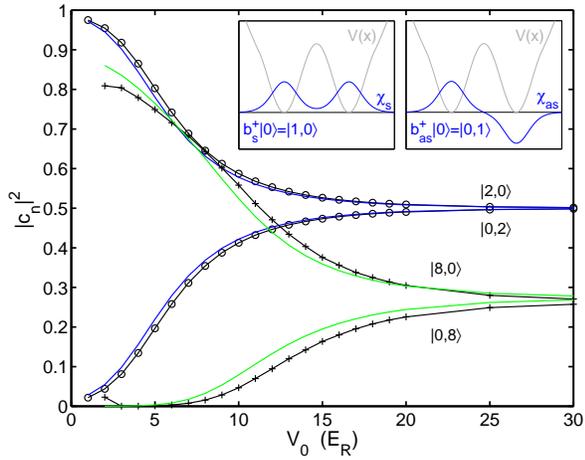} 
\caption{(Color online) The lowest coefficients $c_n$ for two ($\circ$) and eight atoms ($+$) in a double well potential
in dependency on the lattice depth $V_0$ calculated by exact diagonalization.
The lines without markers are obtained using the LBA.
The inset shows a symmetric and an antisymmetric basis wave function.}
\label{Coeff} 
\end{figure}

The short range interaction potential of ultracold bosonic atoms can be approximated by 
a contact potential  $g\delta(\mathbf r -\mathbf r')$ with the interaction parameter 
$g=\frac{4\pi \hbar^2}{m} a_s$, where $a_s$ is the $s$-wave scattering length and 
$m$ the mass of the atoms \cite{Leggett}. 
Thus, the Hamiltonian including the full two-particle interaction in a periodic potential $V_P$ 
is given by 
\begin{equation}
  \hat H=\intvecr \hat\psi^\dagger(\mathbf r) \left[
     \frac{\hat {\mathbf p}^2}{2m}+ V_P(\mathbf r)+ \frac{g}{2} \hat\psi^\dagger(\mathbf r) \hat\psi(\mathbf r)
  \right] \hat\psi(\mathbf r),
  \label{H}
\end{equation}
where $\hat\psi(\mathbf r)$ is the bosonic field operator. 
We use the potential   $V_P=V_{0x}\cos^2(kx)+V_{0y}\cos^2(ky)+V_{0z}\cos^2(kz)$   
that is truncated to $(n_x,1,1)$ sites for chains and $(n_x,n_y,1)$ sites for two-dimensional lattices.
The periodic potential is continued at its boundary by a harmonic confinement potential
\cite{Note1} (see inset of Fig.~\ref{Coeff}).  
We  model  an optical lattice with the periodicity $a=515\text{~nm}$ ($k={\pi}/{a}$)
and vary the depth of the lattice from $1\ER$ to $40\ER$
given in units of the recoil energy $E_R=\frac{\hbar^2 k^2}{2m}=2.16\,h\,\text{kHz}$.
The quasi-one-dimensional chains  and two-dimensional lattices have a 
transversal confinement $V_{0y}=V_{0z}=40\ER$ and $V_{0z}=40\ER$, respectively, 
so that the transversal tunneling can be neglected.
Exact diagonalization is performed  in the Bloch representation of the optical lattice.
By using a few-particle basis the two-particle interaction is fully included.
The truncation of the basis  at a sufficiently high energy allows the inclusion of orbital effects 
\cite{Note2}.
The calculations are performed for $^{87}$Rb atoms which are present in many experimental
setups.

%%%%%%%%%%%%%%%%%%%%%%%%%%%%%%%%%%%%%%%%%%%%%%%%%%%%%%%%%%%%%%%%%%%%%%%%%%%%%%%%%%%%%%%%%%%%%%%%
%%%%%%%%%%%%%%%%%%%%%%%%%%%%%%%%%%%%%%%%%%%%%%%%%%%%%%%%%%%%%%%%%%%%%%%%%%%%%%%%%%%%%%%%%%%%%%%%

\section{A precursor of the Mott insulator in finite systems}

We start the discussion with  atoms in a double well which 
exhibit a crossover reminiscent  
of the superfluid to Mott insulator transition.
 Although the double well with commensurate filling represents an intuitive and easy-to-handle model, 
the eigenstates and the spectrum have 
already a structure similar to the studied chains with 3 to 10 sites.
A double well with two particles  
can easily be treated analytically with the restriction to the lowest band.
 The Hilbert space separates into states with even and odd parity 
that do not couple.
The subspace with even parity comprises of  
the states $\ket{2,0}=\frac{1}{\sqrt{2}} b_{\s}^{\dagger2}\ket{0}$ and $\ket{0,2}=\frac{1}{\sqrt{2}} b_{\as}^{\dagger2}\ket{0}$, where $b_{\s}^{\dagger}$ is the creation operator
of particles in the symmetric state and $b_{\as}^{\dagger}$ in the antisymmetric state (see inset of Fig.~\ref{Coeff}).
Since the energy difference between symmetric and antisymmetric single-particle states $\chi_{\s}$ and $\chi_{as}$
is twice the tunneling energy $t$, the energy of the states $\ket{2,0}$ and $\ket{0,2}$ differs by  
$2\Omega=4t+\Delta$. 
The difference in the interaction energies $\Delta$ equals $g\intvecr [\chi_{\as}^4(\mathbf r) - \chi_{\s}^4(\mathbf r)]$
using the real space representation.  
The off-diagonal matrix element between both states 
 is $I=g \intvecr \chi_{\s}^2(\mathbf r) \chi_{\as}^2(\mathbf r)$.  
Thus, the ground state is given by $\Psi_{0}=\cos\theta \ket{2,0} + \sin\theta \ket{0,2}$, where
\begin{equation}
  \theta=\text{atan}(\frac {\Omega-\sqrt{\Omega^2+I^2}} {I}).
  \label{theta}
\end{equation}
For very shallow lattices,  when $t$ approaches infinity,   $\theta$ vanishes and  
$\Psi_{0,V_0\rightarrow 0}$ equals $\ket{2,0} $, i.e., both particles 
occupy the energetically lower symmetric one-particle state.

In the limit of deep lattices  ($t\rightarrow 0$) 
the difference between the symmetric and antisymmetric wave function $\chi_{\s}^2(x)-\chi_{\as}^2(x)$
vanishes ($\Delta \rightarrow 0$) and $\theta$ approaches $-\pi/4$.
Consequently, the ground state is given by
$\Psi_{0,V_0\rightarrow \infty}=\frac{1}{\sqrt{2}} \ket{2,0} - \frac{1}{\sqrt{2}} \ket{0,2}$.
Using the creation operators of a particle in the left and right well  
$b_{l/r}^{\dagger}=\frac{1}{\sqrt{2}} (b_{\s}^{\dagger}\pm b_{\as}^{\dagger}) $
the ground state can be rewritten as $\Psi_{0,V_0\rightarrow \infty}=b_l^{\dagger} b_r^{\dagger} \ket{0}$
and it becomes obvious that one particle is localized on the left site and one particle on the right site.
In addition to the localization, a fundamental property of a Mott-insulator-like state is an excitation gap
which is given here by $2I$. 
The first excited state with an uneven parity and the second excited state 
$\Psi_{1/2,V_0\rightarrow \infty}=\frac{1}{2}(b_l^{\dagger 2}\mp b_r^{\dagger 2})\ket{0}$
are degenerate \cite{Note3}.
Both states are symmetric-antisymmetric combinations of doubly occupied sites and thus
represent particle-hole excitations.
In chains and lattices these linear combinations of particle-hole excitations build up the excited band
as discussed further below.

\setlength{\textheight}{23.6cm}

\begin{figure}[b] 
\includegraphics[width=0.9\linewidth]{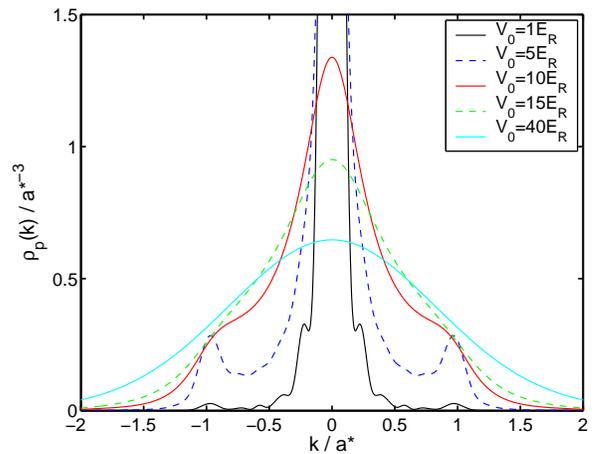} 
\caption{(Color online) The momentum distribution of six bosons in a quasi-one-dimensional chain with six sites. 
The crossover from a delocalized wave function ($V_0\lesssim 5\ER$) to a 
localized wave function ($V_0\gtrsim 10\ER$) can be observed.}
\label{Momentum611}
\end{figure}

Some of the coefficients  $c_n$ in the expansion of the many-body wave function  of the ground state 
are plotted in Fig.~\ref{Coeff}.
Already for a double well it is instructive to compare the
results that  are  obtained using the lowest band approximation (LBA)
with numerical multiband calculations.
For $V_0\lesssim10\ER$ the deviations are well noticeable, whereas
in deep lattices the LBA leads to nearly perfect results for filling factor $\nu=1$. 
For higher filling factors the total interaction energy and 
consequently the deviations increase, 
since higher one-particle bands are occupied in order to minimize the interaction energy.
Exemplarily, the coefficients of the states $\ket{8,0}$ and $\ket{0,8}$ for filling factor $\nu=4$ are plotted in Fig.~\ref{Coeff}. Additionally, the lowest band coefficients $\ket{6,2}$, $\ket{4,4}$, and $\ket{2,6}$ contribute. 
For higher filling factors than $\nu=1$ the deviations do not vanish in deep lattices, 
since in that case $\nu$ interacting particles are trapped on each site, 
leading to a modification of the effective single particle orbitals. 
This has direct implications for the Bose-Hubbard model \cite{Jaksch,HubbardToolbox} which 
is widely used in this context and is commonly restricted to the lowest band. 
Figure~\ref{Coeff} shows that  for  higher  filling factors  ($\nu\gtrsim3$)  this restriction leads to noticeable deviations 
from a multiband calculation, whereas for low filling factors the deviations are small. 
Experimentally, deviations for higher fillings have been observed, e.g., by measuring 
the on-site interaction energy~\cite{Note4,Campbell:313:649}. 

\begin{figure}
\includegraphics[width=0.9\linewidth]{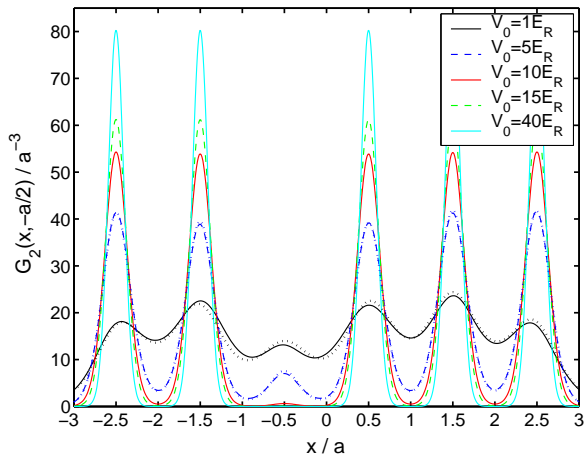} 
\caption{(Color online) The pair correlation function of six bosons in a chain with six sites shows the 
localization of particles for $V_0\gtrsim 10\ER$. 
Due to the low filling factor $\nu=1$, the deviation from results obtained using the LBA (dotted lines) are 
only noticeable for $V_0\lesssim 5\ER$.\\}
\label{Corr611}
\end{figure}

In the following, we extend the double well  to a chain with $N_s=6$ lattice sites and six particles.   
The momentum distribution  $\rho_p(k_x)$ of the chain is shown for different lattice depths $V_0$ in Fig.~\ref{Momentum611}, 
where $k_y=k_z=0$.  
For very shallow lattices ($V_0=1\ER$) a narrow central peak indicates the delocalization
of all particles over the lattice,
whereas for deep lattices a broad Gaussian momentum distribution is observable.
The latter can be assigned to particles that are localized in the center of a single lattice   site.  
Although the system is very small the similarity to macroscopic experimental
results  \cite{Greiner,stoferle:130403}  is striking. 
This is a first indication that the localization mechanism is not very size dependend (see also \cite{Roth}) which
is discussed in detail below.

For $V_0=1\ER$ minor dips in the momentum distribution (Fig.~\ref{Momentum611}) at 
$k_n=(n/N_s)a^*$  are observable, where $a^*=2\pi/a$ is the reciprocal lattice vector. 
These dips originate from the suppression of standing waves in the confinement with odd
parity and wavelengths $\lambda_n=N_s\, a / n$ with $n=1,2,...$,
since the ground state has an even parity.  
Increasing the lattice depth to $V_0=5\ER$, 
 Bragg peaks located at the reciprocal lattice vector appear.  
At the same time, 
the central peak drops rapidly in height and
becomes broader, i.e., the particles begin to separate 
into different wells as the interaction grows relative to the tunneling.  
At $V_0=10\ER$  the minima are smeared out and  
only a small modulation of momentum density due to delocalized particles remains.  
This progress proceeds with increasing lattice depth, so that for approximately $V_0=30\ER$ 
the momentum distribution has a Gaussian shape, corresponding to completely localized particles  
\cite{Note5}.  

\begin{figure}
\includegraphics[width=0.9\linewidth]{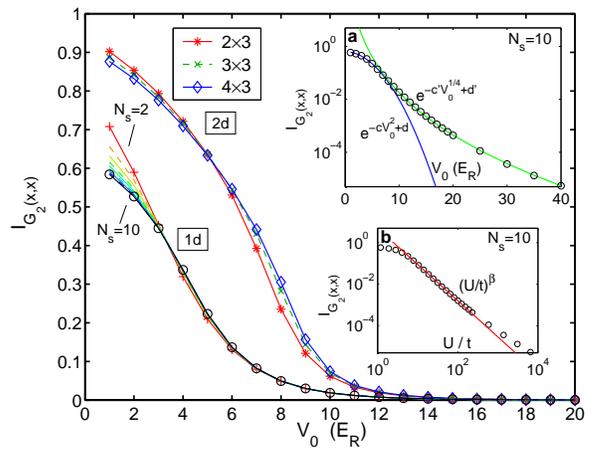}
\vspace{0.3ex} 
\caption{(Color online) The integral over the local correlation function for chains with $N_s$ sites 
and small two-dimensional lattices (6, 9, and 12 sites) with filling factor $\nu=1$ (in LBA). 
Inset (a) is a logarithmic plot in units of $V_0/E_R$ and inset (b) is a double logarithmic plot
in units of $U/J$ for a chain with ten sites.}
\label{UnityCorrInt}
\end{figure}

However, a Gaussian momentum distribution 
may also arise from a superposition of delocalized states and
does not prove the localization on single lattice sites.
Therefore, the pair correlation function 

\begin{equation}
G_2(\mathbf r, \mathbf r')=\frac{1}{\rho(\mathbf r')} \expect{\hat\psi^\dagger(\mathbf r) \hat\psi^\dagger(\mathbf r') \hat\psi(\mathbf r') \hat\psi(\mathbf r)}
\end{equation}
at $y=z=0$  is  studied, which reflects the conditional density of ${N-1}$ particles if one particle can be found at $\mathbf r'$. 
In Fig.~\ref{Corr611} the pair correlation is shown for $x'=-\frac{a}{2}$, i.e., for one particle
located on the third site.
For $V_0\lesssim 5\ER$ other particles can be found on the same site,
whereas for $V_0\gtrsim 10\ER$ the pair correlation vanishes on the third site completely.
Hence, that site and consequently all sites are occupied with exactly one particle in deep lattices. 

The integral over the \textit{local} correlation function 
$I_{G_2(x,x)}=\frac{N/(N-1)}{\intx \rho(x)} \intx G_2(x,x)$
measures the average probability of finding two particles at the same position
and is consequently a good measure for the total spatial correlation of particles.
In Fig.~\ref{UnityCorrInt} the local correlation integral is shown for chains 
with $N_s$ sites and filling factor $\nu=1$.
The calculations are restricted to the lowest band which is quite accurate due to 
the low filling (see Fig.~\ref{Corr611}).

Overall we see an exponential decay of the correlation integral with increasing 
lattice depth.
In deep lattices the integral vanishes which reflects the localization
of all particles.
For $V_0\lesssim 6\ER$ the correlation decreases with an exponent $-cV_0^2$
[see fit for $N_S=10$ in the inset (a) of Fig.~\ref{UnityCorrInt}]. 
Thus, an increase of the potential barriers causes a relatively strong separation of particles
in this regime due to the strong overlap of wave functions.
For $V_0\gtrsim 8\ER$ the particles are located predominately on single lattice sites
and the extent to neighboring sites is small. 
In this region the exponential decay is weaker 
and the correlation drops with an exponent $-c' V_0^{1/4}$.
It is remarkable that our calculations show 
a crossover between two different correlation regimes already for a small sized system.

\setlength{\textheight}{24cm}

In the studied systems we observe a partial loss of spatial correlation in the region
corresponding to the superfluid phase as also found in Ref.~\cite{Roth}.  
Moreover, we see finite correlations above the transition point, 
which can be understood quantitatively by using a perturbative ansatz to first order 
in $t/U$ \cite{Gerbier,spielman:080404}, where $t$ is the tunneling matrix element and
$U$ the on-site interaction energy.  
The perturbed wave function can be written as
$\Psi'=\Psi_{MI}^{\nu=1}+\frac{t}{U}\sum_{ij}b_i^\dagger b_j \Psi_{MI}^{\nu=1}$ where
$\Psi_{MI}^{\nu=1}$ is the pure Mott insulator state at infinite lattice depth 
which has a vanishing local correlation.
The operator $b_i^\dagger b_j$ creates a particle-hole state with a hole on site $j$ 
and a doubly occupied site $i$.
For doubly occupied sites the pair correlation has a constant value 
if neglecting the interaction.
Hence, the expectation value of the correlation integral is roughly 
proportional to $(t/U)^2$.
In a double-logarithmic plot in units of $U/t$ [inset (b) of 
Fig.~\ref{UnityCorrInt}] the correlation integral shows a linear behavior between 
$V_0=6\ER$ and $17\ER$, i.e., the integral is proportional to 
$(U/t)^\beta$.
We observe a value for $\beta$ that is slightly above $\beta=-2$ (about $13\%$).
For shallow lattices the simple perturbative ansatz is 
obviously not suited and for very deep lattices ($V_0>20\ER$) the expectation values are
above the power law fit (see also Ref.~\cite{Gerbier}).  

Only in very shallow lattices the correlation integral varies noticeably with the number
of lattice sites, whereas for $V_0>3\ER$ the integral is nearly size independent
(even for a double well system).
Additionally, the differences for $V_0<3\ER$ are quite small for more than four lattice sites. 
Thus, the localization of particles depends very weakly on the number 
of lattice sites $N_s$, which indicates that the localization may occur  
in the same manner also in chains of macroscopic size. 
We conclude that the blocking mechanism,
which is caused by the tunneling prohibiting repulsion,
is to a large extend insensitive to changes of the system size.
Apparently, the coherence length in the insulating region drops below the extend of the system,
which becomes consequently a good representation of a larger one.  
This explains the similarity of the presented momentum distributions
and experimental results.

\begin{figure}[b]
\includegraphics[width=0.9\linewidth]{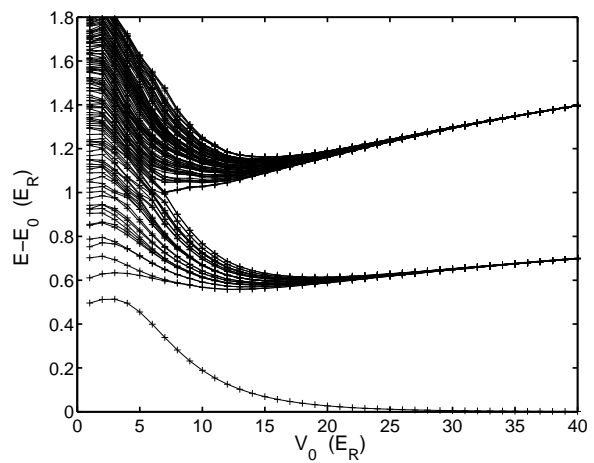} 
\caption{The energy spectrum of six bosons in a chain with six sites relative to the energy $E_0$ 
of the noninteracting system. 
 The spectrum shows the ground state and the two lowest bands.  }
\label{Energies611}
\end{figure}

Exemplarily, the energy spectrum of a system with $N_s=6$ sites, plotted in Fig.~\ref{Energies611}, shows 
the formation of narrow many-particle bands for deep lattices. 
The spectrum is in accordance with the experimentally
observed excitation spectrum in Ref.~\cite{stoferle:130403}.
The  bands are gapped by the interaction energy $U$ of two particles on the same site,
so that, for example, the first excited band consists of states where one particle interacts on average 
with one other particle on the same site.
The number of states in each band is given by the possibilities to remove a certain number
of particles and put them onto other sites resulting in an interaction energy of $nU$
(e.g., the first band has 30 states due to 6 possible sites for a hole
and 5 possible sites with double occupation).
In particular, the energy of the nondegenerate ground state decreases 
and the ground state becomes separated from the excited states which is 
characteristic for  an  incompressible Mott insulating state. 
In the limit of deep lattices the interaction energy of the ground state vanishes, since
the wave function overlap decreases due to the localization on single sites.
The  energy of the  excited  bands slowly increases due to the stronger confinement.
Since the sites are equivalent, the eigenstates of the bands are delocalized.
These delocalized states that form the excited bands in this commensurate system  
reappear within the ground-state band of noncommensurate systems, which are
discussed in the next section.  

\begin{figure}
\includegraphics[width=8.5cm]{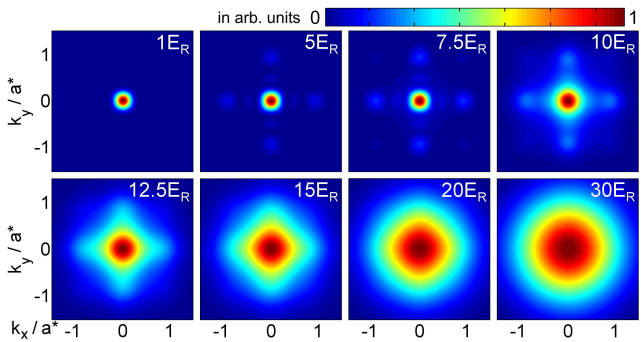} 
\caption{(Color online)  The momentum distribution (normalized to~1) of a two-dimensional $3\times 3$ lattice 
with nine atoms 
for various lattice depths $V_0$ using the LBA.  
}
\label{Momentum2d}
\end{figure}

The results obtained so far for momentum distributions, correlation functions, and energy spectra
of quasi-one-dimensional chains can be generalized to lattices.
As an example, we study a quasi-two-dimensional lattice with $3\times 3$ lattice sites with integer filling factor $\nu=1$.
The momentum distribution is presented in Fig.~\ref{Momentum2d} for different potential depths.
It shows the  crossover  from delocalization to localization
and compares well with experimental results \cite{spielman:080404}.   
For $5\ER$ and $7.5\ER$  Bragg  peaks at $\mathbf k=\pm a^* {\mathbf{\hat k}}_x$  and $\mathbf k=\pm a^* {\mathbf{\hat k}}_y$ 
appear in the distribution.
Due to the finite number of sites per dimension, additional dips at $k=\frac{n}{3}a^*$ can be observed.
Increasing the lattice depth, the momentum distribution smears out due to the localization of particles.
This starts at approximately  $10\ER$ and is far advanced at $12.5\ER$,
which is in accordance with the critical point for infinite systems
\cite{elstner:59:12184}.  
The two-dimensionality of the system becomes apparent for $V_0$ between 
$10\ER$ and $15\ER$.
The momentum distribution reflects the square symmetry of  the boundary of  a single lattice site.
At roughly $V_0=30\ER$ the distribution becomes a Gaussian which
indicates that the particles are located deep in the wells.
Then the confining potential experienced by the particles is circular symmetric and can be approximated
by a two-dimensional harmonic oscillator, 
which is reflected in the momentum distribution.

\setlength{\textheight}{24.2cm}

A more detailed picture of the localization process is given by the integral 
over the local correlation function $I_{G_2(x,x)}$
which is plotted in Fig.~\ref{UnityCorrInt}.
Compared with the quasi-one-dimensional chain, the localization process is shifted 
towards deeper potentials because of two reasons: 
The interaction energy on each lattice site is diminished due to smaller confinement in the $y$ direction and
the tunneling is enhanced since tunneling to four nearest neighbors is possible.
For the two-dimensional lattice, we observe the formation of gapped excited bands in deep lattices as well as 
the separation of the ground state in the energy spectrum in the same way as discussed for chains.

~

We are aware that the finite systems discussed here cannot exhibit 
macroscopic phases and a phase transition.
Nonetheless, the precursors to the superfluid phase and the Mott insulator map
on many aspects known from macroscopic systems. 
The momentum distribution (Figs.~\ref{Momentum611} and \ref{Momentum2d}), 
the correlation function (Figs.~\ref{Corr611} and \ref{UnityCorrInt})
and the energy spectrum (Fig.~\ref{Energies611}) show
that the localization process is in good agreement with infinite systems.
Finite size effects can be observed but do not dominate the behavior of the system.
Therefore, simulations with few lattice sites are also quite applicable to
larger systems and offer an intuitive and detailed insight
due to the accuracy  and the inclusion of  orbital effects.
In the next section we use our method to examine systems 
with noncommensurate filling. 

%%%%%%%%%%%%%%%%%%%%%%%%%%%%%%%%%%%%%%%%%%%%%%%%%%%%%%%%%%%%%%%%%%%%%%%%%%%%%%%%%%%%%%%%%%%%%%%%
%%%%%%%%%%%%%%%%%%%%%%%%%%%%%%%%%%%%%%%%%%%%%%%%%%%%%%%%%%%%%%%%%%%%%%%%%%%%%%%%%%%%%%%%%%%%%%%%

\section{Noncommensurate filling}

When a noncommensurate filling of the lattice is present the physical situation becomes more complicated.  
The localization of all particles as in the Mott-insulator-like regime is suppressed 
by the symmetry of the  potential,  
since the  equivalence of sites  requires the same filling on all sites. 
Consequently, particles which in principle would prefer localization must delocalize over the whole lattice.
The differences to commensurate filling can be illuminated by considering a double well 
with three atoms restricted to the lowest band.   
The basis of even parity states  comprises of  the two states 
$\ket{3,0}=\frac{1}{\sqrt{6}} b_{\s}^{\dagger3}\ket{0}$ and 
$\ket{1,2}=\frac{1}{\sqrt{2}} b_{\s}^{\dagger} b_{\as}^{\dagger2} \ket{0}$. 
The difference in energy between both basis elements is  $2\Omega=4t+\Delta$  
with $\Delta= g\intvecr[\chi_{\as}^4(\mathbf r) + 4 \chi_{\s}^2(\mathbf r) \chi_{\as}^2(\mathbf r) - 3 \chi_{\s}^4(\mathbf r)]$ and 
the off-diagonal matrix element is
$I= \sqrt{3}g \intvecr \chi_{\s}^2(\mathbf r) \chi_{\as}^2(\mathbf r)$. 
Thus, the solution for the ground state is given by the same expression as for two particles, i.e., 
$\Psi_{0}=\cos\theta \ket{3,0} + \sin\theta \ket{1,2}$. 
In deep lattices the tunneling energy  $t$  vanishes, but $\Delta$, which had vanished for two particles,
becomes $4I/\sqrt{3}$.

\begin{figure}[t]
\includegraphics[width=0.9\linewidth]{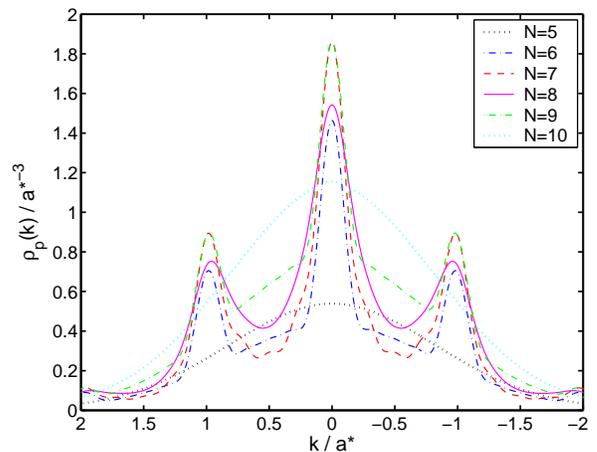} 
\caption{(Color online) The momentum distribution of $N=5$ to 10 atoms in a chain with five sites at $V_0=40\ER$. 
For a perfectly symmetric potential the localization of all particles can only be observed at integer filling
(dotted lines).}
\label{511Momentum}
\end{figure}

The ground state in the limit of deep lattices is given by 
$\Psi_{0,V_0\rightarrow \infty}=\frac{\sqrt{3}}{2} \ket{3,0} - \frac{1}{2} \ket{1,2}=
\frac{1}{2} b_l^{\dagger} b_r^{\dagger} (b_l^{\dagger}+b_r^{\dagger}) \ket{0} $,
since $\theta$ becomes $2{\pi}/{3}$.
This represents a wave function with two atoms that are localized and one atom that is delocalized between both wells. 
Consequently, the ground state is a mixture of localized and delocalized particles.
Of great importance in this context is that in this limit the first excited state, 
which has odd parity and is given by $\Psi_{1,V_0\rightarrow \infty}=
\frac{1}{2} b_l^{\dagger} b_r^{\dagger} (b_l^{\dagger}-b_r^{\dagger}) \ket{0} $, becomes degenerate with
the ground state. 
Small asymmetries lead to linear combinations of the two quasidegenerate states 
and result in
the nonsymmetric states $\frac{1}{\sqrt{2}} b_l^{\dagger2} b_r^{\dagger} \ket{0}$ and  
$\frac{1}{\sqrt{2}} b_l^{\dagger} b_r^{\dagger2} \ket{0}$.
Consequently, in a potential with broken symmetry  
the \textit{third} particle can localize in one of the wells, if the lattice is deep enough.
This localization process depends on the potential difference between the two wells
compared to the energy difference between ground and first excited state $E_1(V_0)-E_0(V_0)$.
The third and fourth excited state which form the "first excited band"
are separated from the ground state by $4I/\sqrt{3}$. 

These intuitive results for a double well  transfer nicely to  chains and even two-dimensional lattices.
Exemplarily, a chain with $N_s=5$ sites filled by 5 to 10 particles is studied.
The momentum distribution
for deep lattices ($V_0=40\ER$) obtained by exact diagonalization is shown in Fig.~\ref{511Momentum}.
For integer filling factors ($N=5$ and $N=10$) a Mott insulator momentum distribution
can be observed as discussed in the previous section.
When, for example, adding a sixth particle to a chain with filling factor $\nu=1$, 
the additional particle cannot localize, since all lattice sites are equivalent.
Despite being delocalized at different sites, the particle has a high probability density
at the lattice site centers.
Therefore, the momentum distribution reflects the lattice structure very clearly.
Correspondingly, it shows peaks at $0$ and $\pm a^*$ as well as smaller peaks at 
$\pm n a^*$ with $n=2,3,...$, which originate from the delocalized particles. 
Additional to this peak structure the momentum distribution has an underlying Gaussian background,
which arises from localized particles.
Recapitulating the ground state for three particles in a double well 
$\Psi_{0,V_0\rightarrow \infty}=\frac{1}{2} b_l^{\dagger} b_r^{\dagger} (b_l^{\dagger}+b_r^{\dagger}) \ket{0} $,
the interpretation is straight forward:
For noninteger filling factors $\nu$, the number of particles may be written as 
$N=\kappa N_s+N_\text{add}$ with the corresponding integer filling factor $\kappa$ 
and the number of additional particles $N_\text{add}$.
In deep lattices $\kappa N_s$ particles localize in the wells of the lattice and the remaining 
$N_\text{add}$ are delocalized. 

The plotted momentum distribution for $V_0=40\ER$ in Fig.~\ref{511Momentum} shows that for
$N=6$ particles the Gaussian background is noticeably smaller than for five particles.
This indicates that the localized particles are influenced by
the delocalized hopping particle which 
experiences the same repulsive interaction on all lattice sites
and consequently interacts with all localized particles.
For seven particles the height of the peaks increases due to two hopping particles, whereas
for eight particles the background increases, since more sites are doubly occupied.
Finally, for nine particles all sites except one are doubly occupied which is equivalent to the
tunneling of a hole in a lattice with filling $\nu=2$.
We also find the delocalization for noncommensurate filling  
by analyzing the pair correlation function $G_2(x,x')$ \cite{Note6}.  

\begin{figure}
\includegraphics[width=0.9\linewidth]{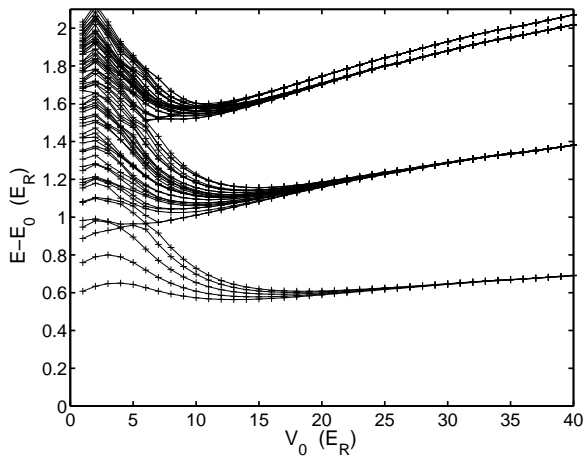} 
\caption{The energy spectrum for six particles on five sites shows the formation of a lowest many-particle band.}
\label{511Energy6}
\end{figure}

The energy spectrum for six particles in a chain with $N_s=5$ sites is shown 
in Fig.~\ref{511Energy6}. 
For deep lattices the formation of bands can be observed, which are roughly separated by the
two-particle interaction energy $U$.
An interesting feature is the splitting of the second excited band which would not 
be observable in the LBA. 
The states with higher energy have three doubly occupied sites whereas the other states 
have one triply occupied site.
Their energy is reduced by a stronger deformation of the wave function
accounting for the effective repulsive potential created by the two other atoms at the same site.
As seen before for a double well, the ground state becomes quasidegenerate 
and lies within a band of $N_s$ states.
Therefore, the ground state is extremely sensitive to small perturbations
of the lattice potential,  which is discussed in the next section.   

\begin{figure}
\includegraphics[width=8.5cm]{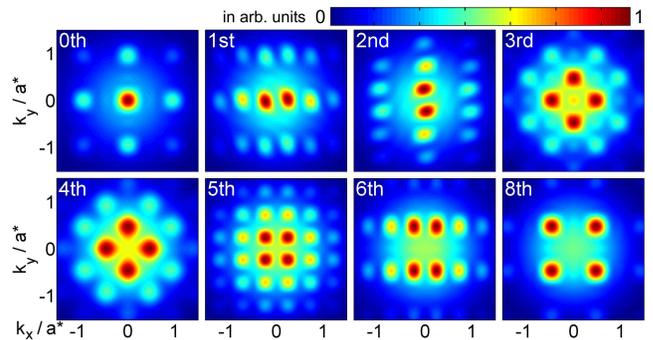} 
\caption{(Color online)  The momentum distribution (normalized to~1) of the lowest band states  of a two-dimensional $3\times 3$ lattice with ten atoms at $V_0=30\ER$ (in LBA).
The seventh excited state, which corresponds to the sixth excited state (rotated by $\pi/2$), is not shown here
[29]. 
}
\label{Momentum2d-10}
\end{figure}

It is hardly suprising that the rich physics of noncommensurate filling 
can also be found in two-dimensional lattices, including the formation of a lowest band 
with its discussed implications.
For a two-dimensional lattice it is interesting to explore the  quantum mechanical nature of the states 
that are contributing to the lowest band. 
For a $3\times 3$ lattice with ten particles the lowest band consists of nine states
due to the one additional particle in comparison with commensurate filling. 
Similar to Fig.~\ref{511Energy6} these states become quasi-degenerate with increasing lattice depth
representing all delocalized combinations with the 
two-particle interaction energy $U$.
In Fig.~\ref{Momentum2d-10} the momentum distribution of the lowest band states is plotted at $V_0=30\ER$
(the seventh state is not shown) \cite{Note7}.   
The strong interference pattern of the momentum distribution is in eye-catching contrast with
the Gaussian shape of a Mott insulator state and reflects the delocalization.   
The general structure of the shown band states  with only few states can be generalized to larger 
noncommensurately filled lattices and to particle-hole excitations for commensurate filling.

%%%%%%%%%%%%%%%%%%%%%%%%%%%%%%%%%%%%%%%%%%%%%%%%%%%%%%%%%%%%%%%%%%%%%%%%%%%%%%%%%%%%%%%%%%%%%%%%
%%%%%%%%%%%%%%%%%%%%%%%%%%%%%%%%%%%%%%%%%%%%%%%%%%%%%%%%%%%%%%%%%%%%%%%%%%%%%%%%%%%%%%%%%%%%%%%%

\section{Noncommensurate filling in a harmonic confinement}

\begin{figure}
\includegraphics[width=0.9\linewidth]{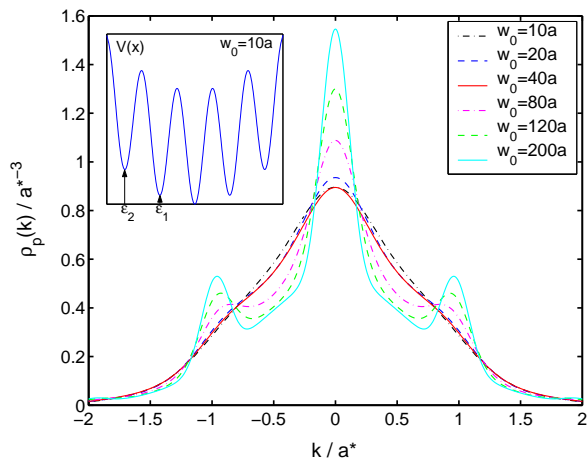} 
\caption{(Color online) The momentum distribution of six atoms in a chain with five sites at $V_0=20\ER$ 
for a laser beam with a finite Gauss width $w_0$.
The inset shows the corresponding site offsets $\epsilon_1$ and $\epsilon_2$.}
\label{511MomentumGauss}
\end{figure}

As a last point we discuss the noncommensurate filling under the influence of an additional 
symmetry-breaking potential. 
Perturbations of the potential can cause linear combinations of the lowest band states  
that allow  the  localization of the additional particles on specific sites. 
Consequently, a small external confinement can destroy the partly delocalized phase 
in the same way  as  small random site offsets caused by lattice fluctuations  
in a Bose glass \cite{Fischer}. 
The parameter which triggers the localization of the additional $N_\text{add}$ particles is the bandwidth
of the lowest band which must be similar or smaller than the site offsets to attain localized particles.
In experimental setups perfectly flat potentials are hard to achieve, due to the finite
waist of the laser beams, which establish the optical lattice, and additional external fields. 
We investigate this effect by using a chain with five sites and six particles that experience 
an additional potential $-2V_T e^{-2x^2/w_0^2}$ with $V_T=40\ER$.
The potential is motivated by the 
Gaussian beam waist $w_0$ given in units of the lattice constant $a$,
but it can also be approximated  
by the harmonic potential $\frac{4V_T}{w_0^2}x^2$.
The momentum distribution at $V_0=20\ER$ is shown in Fig.~\ref{511MomentumGauss} for different 
beam widths $w_0$ ranging from $w_0=10 a$ to $w_0=200 a$.
The corresponding offset energies relative to the central site are given by
$\epsilon_1=\frac{160}{w_0^2/a^2}\ER$ and $\epsilon_2=4\epsilon_1$ (see inset of Fig.~\ref{511MomentumGauss}).
The width of the ground state band at $V_0=20\ER$ is roughly $\frac{1}{50}\ER$.

In order to localize the additional particle, the offset $\epsilon_1$ 
is important with respect to the bandwidth.
For $w_0=200a$ and $w_0=120a$ peaks due to delocalization can be well identified in the
momentum distribution, 
but at $w_0=80a$ ($\epsilon_1=\frac{1}{40}\ER$) this structure is smeared out.
At $w_0=40a$ ($\epsilon_1=\frac{1}{10}\ER$) the momentum distribution matches 
the distribution for commensurate filling shown in Fig.~\ref{Momentum611}  (at $V_0\approx 15\ER$),  
i.e., the sixth particle is localized in the center of the lattice.
Below $w_0=40a$ the momentum distribution does not change noticeably. 
Instead, the {\it density} changes drastically, since the gain in potential energy $\epsilon_2$
exceeds the repulsive interaction of two particles at the same site $U$.
At $w_0=20a$ the density at the two outer sites vanishes and the inner sites are doubly occupied,
due to a total energy gain of roughly $ 2(\epsilon_2-U) \approx 2\times (1.6\ER-0.6\ER$).
For the stronger confinement $w_0=10a$ the central site is even occupied by four particles. 

Despite the harmonic potential the lattice symmetry can be broken by 
applying different moderate perturbation potentials (such as single site offsets or a linear potential).
The localization process remains qualitatively the same for the studied finite systems, since basicly
local site offsets are responsible for the localization.  
We conclude that in perturbed lattices a Bose-glass-like localization of all particles can 
always be achieved in sufficiently deep lattices triggered by the ratio of 
offset energies and bandwidth.
However, in intermediate lattices a mixed phase with 
$N_\text{add}$ delocalized  particles can be observed, if the lattice fluctuations are smaller than the bandwidth.
Increasing the lattice depth (decreasing the bandwidth) first $\kappa N_s$ particles and
in deeper potentials the remaining $N_\text{add}$ particles localize.
Experimentally the observation of the mixed phase may be hindered by the finite temperature of the BEC.
We note that in stronger harmonic confinements in which the offset energies match the interaction energy $U$ 
($w_0 \lesssim 20a$  for the system above)   a precursor of a  shell structure  can be observed,  
which shows regions with different  occupations  per site \cite{Batrouni}.

%%%%%%%%%%%%%%%%%%%%%%%%%%%%%%%%%%%%%%%%%%%%%%%%%%%%%%%%%%%%%%%%%%%%%%%%%%%%%%%%%%%%%%%%%%%%%%%%
%%%%%%%%%%%%%%%%%%%%%%%%%%%%%%%%%%%%%%%%%%%%%%%%%%%%%%%%%%%%%%%%%%%%%%%%%%%%%%%%%%%%%%%%%%%%%%%%
\vspace{4ex}

\section{Conclusions}

We have studied  bosonic atoms in finite optical lattices using an exact treatment 
which includes effects of higher bands.  
Due to  the equivalence of sites,  finite lattices with integer filling factors 
exhibit a fundamentally different behavior
than those with noninteger filling factors. 
The well studied  superfluid to Mott insulator transition can be
recovered in finite commensurately filled chains and two-dimensional lattices with few lattice sites.
Our results show the localization of atoms in deep lattices and
reveal a striking similarity to the momentum distribution  observed in macroscopic systems.  
Furthermore, we have shown  that the local correlation is widely independent of the system size 
which indicates that the localization process in small systems 
compares with that in infinite systems.  
The localization is also reflected by the formation of an energy gap in the energy spectrum.

For noninteger filling factors  
only  the particles that correspond to integer filling  localize in deep lattices whereas 
the additional particles are delocalized.
The coexistence of localized and delocalized particles in the ground state
can be observed in the momentum distribution and the pair correlation function.
The energy spectrum shows the formation of a narrow lowest band
in deep lattices causing the ground state to be extremely sensitive to 
perturbations of the potential 
such as lattice imperfections or additional confinements.
Triggered by the ratio of bandwidth and 
site offsets one observes the localization of all particles
which is similar to the localization process  in a Bose glass.   
In weakly confined systems this leads to a localization which
occurs in deeper potentials than in lattices with commensurate filling.

Briefly, we have shown how the macroscopic physics of the Mott insulator
and of the Bose glass transfer to finite systems and
that the detailed simulation of small systems offer important information about larger ones.
Especially, we gained detailed insight into the localization behavior of 
experimentally relevant finite systems.

\section{Acknowledgments}

We thank Frank Deuretzbacher for valuable discussions.

%%%%%%%%%%%%%%%%%%%%%%%%%%%%%%%%%%%%%%%%%%%%%%%%%%%%%%%%%%%%%%%%%%%%%%%%%%%%%%%%%%%%%%%%%%%%%%%%
%%%%%%%%%%%%%%%%%%%%%%%%%%%%%%%%%%%%%%%%%%%%%%%%%%%%%%%%%%%%%%%%%%%%%%%%%%%%%%%%%%%%%%%%%%%%%%%%


\begin{thebibliography}{30}
\expandafter\ifx\csname natexlab\endcsname\relax\def\natexlab#1{#1}\fi
\expandafter\ifx\csname bibnamefont\endcsname\relax
  \def\bibnamefont#1{#1}\fi
\expandafter\ifx\csname bibfnamefont\endcsname\relax
  \def\bibfnamefont#1{#1}\fi
\expandafter\ifx\csname citenamefont\endcsname\relax
  \def\citenamefont#1{#1}\fi
\expandafter\ifx\csname url\endcsname\relax
  \def\url#1{\texttt{#1}}\fi
\expandafter\ifx\csname urlprefix\endcsname\relax\def\urlprefix{URL }\fi
\providecommand{\bibinfo}[2]{#2}
\providecommand{\eprint}[2][]{\url{#2}}

\bibitem[{\citenamefont{Jessen and Deutsch}(1996)}]{JessenReview}
\bibinfo{author}{\bibfnamefont{P.}~\bibnamefont{Jessen}} \bibnamefont{and}
  \bibinfo{author}{\bibfnamefont{I.}~\bibnamefont{Deutsch}},
  \bibinfo{journal}{Adv. Atom., Mol., Opt. Phys.}
  \textbf{\bibinfo{volume}{37}}, \bibinfo{pages}{95} (\bibinfo{year}{1996}).

\bibitem[{\citenamefont{{Inouye} et~al.}(1998)\citenamefont{{Inouye},
  {Andrews}, {Stenger}, {Miesner}, {Stamper-Kurn}, and {Ketterle}}}]{Inouye}
\bibinfo{author}{\bibfnamefont{S.}~\bibnamefont{{Inouye}}},
  \bibinfo{author}{\bibfnamefont{M.~R.} \bibnamefont{{Andrews}}},
  \bibinfo{author}{\bibfnamefont{J.}~\bibnamefont{{Stenger}}},
  \bibinfo{author}{\bibfnamefont{H.-J.} \bibnamefont{{Miesner}}},
  \bibinfo{author}{\bibfnamefont{D.~M.} \bibnamefont{{Stamper-Kurn}}},
  \bibnamefont{and}
  \bibinfo{author}{\bibfnamefont{W.}~\bibnamefont{{Ketterle}}},
  \bibinfo{journal}{\nat} \textbf{\bibinfo{volume}{392}}, \bibinfo{pages}{151}
  (\bibinfo{year}{1998}).

\bibitem[{\citenamefont{Jaksch et~al.}(1999)\citenamefont{Jaksch, Briegel,
  Cirac, Gardiner, and Zoller}}]{JakschGate}
\bibinfo{author}{\bibfnamefont{D.}~\bibnamefont{Jaksch}},
  \bibinfo{author}{\bibfnamefont{H.-J.} \bibnamefont{Briegel}},
  \bibinfo{author}{\bibfnamefont{J.~I.} \bibnamefont{Cirac}},
  \bibinfo{author}{\bibfnamefont{C.~W.} \bibnamefont{Gardiner}},
  \bibnamefont{and} \bibinfo{author}{\bibfnamefont{P.}~\bibnamefont{Zoller}},
  \bibinfo{journal}{Phys. Rev. Lett.} \textbf{\bibinfo{volume}{82}},
  \bibinfo{pages}{1975} (\bibinfo{year}{1999}).

\bibitem[{\citenamefont{Brennen et~al.}(1999)\citenamefont{Brennen, Caves,
  Jessen, and Deutsch}}]{BrennenGate}
\bibinfo{author}{\bibfnamefont{G.~K.} \bibnamefont{Brennen}},
  \bibinfo{author}{\bibfnamefont{C.~M.} \bibnamefont{Caves}},
  \bibinfo{author}{\bibfnamefont{P.~S.} \bibnamefont{Jessen}},
  \bibnamefont{and} \bibinfo{author}{\bibfnamefont{I.~H.}
  \bibnamefont{Deutsch}}, \bibinfo{journal}{Phys. Rev. Lett.}
  \textbf{\bibinfo{volume}{82}}, \bibinfo{pages}{1060} (\bibinfo{year}{1999}).

\bibitem[{\citenamefont{Schrader et~al.}(2004)\citenamefont{Schrader, Dotsenko,
  Khudaverdyan, Miroshnychenko, Rauschenbeutel, and Meschede}}]{Schrader}
\bibinfo{author}{\bibfnamefont{D.}~\bibnamefont{Schrader}},
  \bibinfo{author}{\bibfnamefont{I.}~\bibnamefont{Dotsenko}},
  \bibinfo{author}{\bibfnamefont{M.}~\bibnamefont{Khudaverdyan}},
  \bibinfo{author}{\bibfnamefont{Y.}~\bibnamefont{Miroshnychenko}},
  \bibinfo{author}{\bibfnamefont{A.}~\bibnamefont{Rauschenbeutel}},
  \bibnamefont{and} \bibinfo{author}{\bibfnamefont{D.}~\bibnamefont{Meschede}},
  \bibinfo{journal}{Phys. Rev. Lett.} \textbf{\bibinfo{volume}{93}},
  \bibinfo{eid}{150501} (\bibinfo{year}{2004}).

\bibitem[{\citenamefont{{Miroshnychenko}
  et~al.}(2006)\citenamefont{{Miroshnychenko}, {Alt}, {Dotsenko},
  {F{\"o}rster}, {Khudaverdyan}, {Meschede}, {Schrader}, and
  {Rauschenbeutel}}}]{Miroshnychenko}
\bibinfo{author}{\bibfnamefont{Y.}~\bibnamefont{{Miroshnychenko}}},
  \bibinfo{author}{\bibfnamefont{W.}~\bibnamefont{{Alt}}},
  \bibinfo{author}{\bibfnamefont{I.}~\bibnamefont{{Dotsenko}}},
  \bibinfo{author}{\bibfnamefont{L.}~\bibnamefont{{F{\"o}rster}}},
  \bibinfo{author}{\bibfnamefont{M.}~\bibnamefont{{Khudaverdyan}}},
  \bibinfo{author}{\bibfnamefont{D.}~\bibnamefont{{Meschede}}},
  \bibinfo{author}{\bibfnamefont{D.}~\bibnamefont{{Schrader}}},
  \bibnamefont{and}
  \bibinfo{author}{\bibfnamefont{A.}~\bibnamefont{{Rauschenbeutel}}},
  \bibinfo{journal}{\nat} \textbf{\bibinfo{volume}{442}}, \bibinfo{pages}{151}
  (\bibinfo{year}{2006}).

\bibitem[{\citenamefont{Sebby-Strabley
  et~al.}(2006)\citenamefont{Sebby-Strabley, Anderlini, Jessen, and
  Porto}}]{Sebby}
\bibinfo{author}{\bibfnamefont{J.}~\bibnamefont{Sebby-Strabley}},
  \bibinfo{author}{\bibfnamefont{M.}~\bibnamefont{Anderlini}},
  \bibinfo{author}{\bibfnamefont{P.~S.} \bibnamefont{Jessen}},
  \bibnamefont{and} \bibinfo{author}{\bibfnamefont{J.~V.} \bibnamefont{Porto}},
  \bibinfo{journal}{Phys. Rev. A} \textbf{\bibinfo{volume}{73}},
  \bibinfo{eid}{033605} (\bibinfo{year}{2006}).

\bibitem[{\citenamefont{Fisher et~al.}(1989)\citenamefont{Fisher, Weichman,
  Grinstein, and Fisher}}]{Fischer}
\bibinfo{author}{\bibfnamefont{M.~P.~A.} \bibnamefont{Fisher}},
  \bibinfo{author}{\bibfnamefont{P.~B.} \bibnamefont{Weichman}},
  \bibinfo{author}{\bibfnamefont{G.}~\bibnamefont{Grinstein}},
  \bibnamefont{and} \bibinfo{author}{\bibfnamefont{D.~S.}
  \bibnamefont{Fisher}}, \bibinfo{journal}{Phys. Rev. B}
  \textbf{\bibinfo{volume}{40}}, \bibinfo{pages}{546} (\bibinfo{year}{1989}).

\bibitem[{\citenamefont{Jaksch et~al.}(1998)\citenamefont{Jaksch, Bruder,
  Cirac, Gardiner, and Zoller}}]{Jaksch}
\bibinfo{author}{\bibfnamefont{D.}~\bibnamefont{Jaksch}},
  \bibinfo{author}{\bibfnamefont{C.}~\bibnamefont{Bruder}},
  \bibinfo{author}{\bibfnamefont{J.~I.} \bibnamefont{Cirac}},
  \bibinfo{author}{\bibfnamefont{C.~W.} \bibnamefont{Gardiner}},
  \bibnamefont{and} \bibinfo{author}{\bibfnamefont{P.}~\bibnamefont{Zoller}},
  \bibinfo{journal}{Phys. Rev. Lett.} \textbf{\bibinfo{volume}{81}},
  \bibinfo{pages}{3108} (\bibinfo{year}{1998}).

\bibitem[{\citenamefont{{Greiner} et~al.}(2002)\citenamefont{{Greiner},
  {Mandel}, {Esslinger}, {H{\"a}nsch}, and {Bloch}}}]{Greiner}
\bibinfo{author}{\bibfnamefont{M.}~\bibnamefont{{Greiner}}},
  \bibinfo{author}{\bibfnamefont{O.}~\bibnamefont{{Mandel}}},
  \bibinfo{author}{\bibfnamefont{T.}~\bibnamefont{{Esslinger}}},
  \bibinfo{author}{\bibfnamefont{T.~W.} \bibnamefont{{H{\"a}nsch}}},
  \bibnamefont{and} \bibinfo{author}{\bibfnamefont{I.}~\bibnamefont{{Bloch}}},
  \bibinfo{journal}{\nat} \textbf{\bibinfo{volume}{415}}, \bibinfo{pages}{39}
  (\bibinfo{year}{2002}).

\bibitem[{\citenamefont{{Gerbier} et~al.}(2005)\citenamefont{{Gerbier},
  {Widera}, {F{\"o}lling}, {Mandel}, {Gericke}, and {Bloch}}}]{Gerbier}
\bibinfo{author}{\bibfnamefont{F.}~\bibnamefont{{Gerbier}}},
  \bibinfo{author}{\bibfnamefont{A.}~\bibnamefont{{Widera}}},
  \bibinfo{author}{\bibfnamefont{S.}~\bibnamefont{{F{\"o}lling}}},
  \bibinfo{author}{\bibfnamefont{O.}~\bibnamefont{{Mandel}}},
  \bibinfo{author}{\bibfnamefont{T.}~\bibnamefont{{Gericke}}},
  \bibnamefont{and} \bibinfo{author}{\bibfnamefont{I.}~\bibnamefont{{Bloch}}},
  \bibinfo{journal}{Phys. Rev. Lett.} \textbf{\bibinfo{volume}{95}},
  \bibinfo{pages}{050404} (\bibinfo{year}{2005}).

\bibitem[{\citenamefont{St{\"o}ferle et~al.}(2004)\citenamefont{St{\"o}ferle,
  Moritz, Schori, K{\"o}hl, and Esslinger}}]{stoferle:130403}
\bibinfo{author}{\bibfnamefont{T.}~\bibnamefont{St{\"o}ferle}},
  \bibinfo{author}{\bibfnamefont{H.}~\bibnamefont{Moritz}},
  \bibinfo{author}{\bibfnamefont{C.}~\bibnamefont{Schori}},
  \bibinfo{author}{\bibfnamefont{M.}~\bibnamefont{K{\"o}hl}}, \bibnamefont{and}
  \bibinfo{author}{\bibfnamefont{T.}~\bibnamefont{Esslinger}},
  \bibinfo{journal}{Phys. Rev. Lett.} \textbf{\bibinfo{volume}{92}},
  \bibinfo{eid}{130403} (\bibinfo{year}{2004}).

\bibitem[{\citenamefont{Damski et~al.}(2003)\citenamefont{Damski, Zakrzewski,
  Santos, Zoller, and Lewenstein}}]{Damski}
\bibinfo{author}{\bibfnamefont{B.}~\bibnamefont{Damski}},
  \bibinfo{author}{\bibfnamefont{J.}~\bibnamefont{Zakrzewski}},
  \bibinfo{author}{\bibfnamefont{L.}~\bibnamefont{Santos}},
  \bibinfo{author}{\bibfnamefont{P.}~\bibnamefont{Zoller}}, \bibnamefont{and}
  \bibinfo{author}{\bibfnamefont{M.}~\bibnamefont{Lewenstein}},
  \bibinfo{journal}{Phys. Rev. Lett.} \textbf{\bibinfo{volume}{91}},
  \bibinfo{pages}{080403} (\bibinfo{year}{2003}).

\bibitem[{\citenamefont{Fallani et~al.}(2007)\citenamefont{Fallani, Lye,
  Guarrera, Fort, and Inguscio}}]{Fallani}
\bibinfo{author}{\bibfnamefont{L.}~\bibnamefont{Fallani}},
  \bibinfo{author}{\bibfnamefont{J.~E.} \bibnamefont{Lye}},
  \bibinfo{author}{\bibfnamefont{V.}~\bibnamefont{Guarrera}},
  \bibinfo{author}{\bibfnamefont{C.}~\bibnamefont{Fort}}, \bibnamefont{and}
  \bibinfo{author}{\bibfnamefont{M.}~\bibnamefont{Inguscio}},
  \bibinfo{journal}{Phys. Rev. Lett.} \textbf{\bibinfo{volume}{98}},
  \bibinfo{eid}{130404} (\bibinfo{year}{2007}).

\bibitem[{\citenamefont{Pugatch et~al.}(2006)\citenamefont{Pugatch, Bar-gill,
  Katz, Rowen, and Davidson}}]{Pugatch}
\bibinfo{author}{\bibfnamefont{R.}~\bibnamefont{Pugatch}},
  \bibinfo{author}{\bibfnamefont{N.}~\bibnamefont{Bar-gill}},
  \bibinfo{author}{\bibfnamefont{N.}~\bibnamefont{Katz}},
  \bibinfo{author}{\bibfnamefont{E.}~\bibnamefont{Rowen}}, \bibnamefont{and}
  \bibinfo{author}{\bibfnamefont{N.}~\bibnamefont{Davidson}},
  \bibinfo{journal}{e-print arXiv:cond-mat/0603571}  (\bibinfo{year}{2006}).

\bibitem[{\citenamefont{Scalettar et~al.}(1991)\citenamefont{Scalettar,
  Batrouni, and Zimanyi}}]{scalettar:3144}
\bibinfo{author}{\bibfnamefont{R.~T.} \bibnamefont{Scalettar}},
  \bibinfo{author}{\bibfnamefont{G.~G.} \bibnamefont{Batrouni}},
  \bibnamefont{and} \bibinfo{author}{\bibfnamefont{G.~T.}
  \bibnamefont{Zimanyi}}, \bibinfo{journal}{Phys. Rev. Lett.}
  \textbf{\bibinfo{volume}{66}}, \bibinfo{pages}{3144} (\bibinfo{year}{1991}).

\bibitem[{\citenamefont{Leggett}(2001)}]{Leggett}
\bibinfo{author}{\bibfnamefont{A.~J.} \bibnamefont{Leggett}},
  \bibinfo{journal}{Rev. Mod. Phys.} \textbf{\bibinfo{volume}{73}},
  \bibinfo{pages}{307} (\bibinfo{year}{2001}).

\bibitem[{Not({\natexlab{a}})}]{Note1}
\bibinfo{note}{The confining potential $V(x)=\gamma V_{0x} \big(
  \frac{2kx}{n\pi}\big)^2$ continues the periodic potential at $|x|=x_0$, where
  $n$ is the number of sites. The potential is continuously differentiable
  which determines $\gamma$ ($\approx 1$) and $x_0$ ($\approx na/2$).}

\bibitem[{Not({\natexlab{b}})}]{Note2}
\bibinfo{note}{We used up to $100\,000$ basis states and parity conservation if
  applicable including the formed Bloch bands and the lowest "continuum" states
  (bound states in the confinement).}

\bibitem[{Not({\natexlab{c}})}]{Note3}
\bibinfo{note}{The first excited state
  $\Psi_{1}=\ket{1,1}=b_{\s}^{\dagger}b_{\as}^{\dagger}\ket{0}$ remains
  unchanged and the second excited state is given by $\Psi_{2}=\cos\theta'
  \ket{2,0} + \sin\theta' \ket{0,2}$ with
  $\theta'=\text{atan}[(\Omega+\sqrt{\Omega^2+I^2})/I]$.}

\bibitem[{\citenamefont{{Jaksch} and {Zoller}}(2005)}]{HubbardToolbox}
\bibinfo{author}{\bibfnamefont{D.}~\bibnamefont{{Jaksch}}} \bibnamefont{and}
  \bibinfo{author}{\bibfnamefont{P.}~\bibnamefont{{Zoller}}},
  \bibinfo{journal}{Ann. Phys. (N.Y.)} \textbf{\bibinfo{volume}{315}},
  \bibinfo{pages}{52} (\bibinfo{year}{2005}).

\bibitem[{Not({\natexlab{d}})}]{Note4}
\bibinfo{note}{We can confirm the experimental observation and the calculation
  in Ref.~\cite{Campbell:313:649} also by the diagonalization of several bosons
  on a single site. The on-site energy $U$ of filling factor $\nu=5$ is, e.g.,
  roughly $17\%$ smaller than for $\nu=1$ at $V_0=30\ER$.}

\bibitem[{\citenamefont{{Campbell} et~al.}(2006)\citenamefont{{Campbell},
  {Mun}, {Boyd}, {Medley}, {Leanhardt}, {Marcassa}, {Pritchard}, and
  {Ketterle}}}]{Campbell:313:649}
\bibinfo{author}{\bibfnamefont{G.~K.} \bibnamefont{{Campbell}}},
  \bibinfo{author}{\bibfnamefont{J.}~\bibnamefont{{Mun}}},
  \bibinfo{author}{\bibfnamefont{M.}~\bibnamefont{{Boyd}}},
  \bibinfo{author}{\bibfnamefont{P.}~\bibnamefont{{Medley}}},
  \bibinfo{author}{\bibfnamefont{A.~E.} \bibnamefont{{Leanhardt}}},
  \bibinfo{author}{\bibfnamefont{L.~G.} \bibnamefont{{Marcassa}}},
  \bibinfo{author}{\bibfnamefont{D.~E.} \bibnamefont{{Pritchard}}},
  \bibnamefont{and}
  \bibinfo{author}{\bibfnamefont{W.}~\bibnamefont{{Ketterle}}},
  \bibinfo{journal}{Science} \textbf{\bibinfo{volume}{313}},
  \bibinfo{pages}{649} (\bibinfo{year}{2006}).

\bibitem[{\citenamefont{Roth and Burnett}(2003)}]{Roth}
\bibinfo{author}{\bibfnamefont{R.}~\bibnamefont{Roth}} \bibnamefont{and}
  \bibinfo{author}{\bibfnamefont{K.}~\bibnamefont{Burnett}},
  \bibinfo{journal}{Phys. Rev. A} \textbf{\bibinfo{volume}{67}},
  \bibinfo{pages}{031602(R)} (\bibinfo{year}{2003}).

\bibitem[{Not({\natexlab{e}})}]{Note5}
\bibinfo{note}{The density on one site corresponds to the density of a single
  particle in a single-sited $\cos^2$ potential, but deviates slightly from the
  harmonic approach.}

\bibitem[{\citenamefont{Spielman et~al.}(2007)\citenamefont{Spielman, Phillips,
  and Porto}}]{spielman:080404}
\bibinfo{author}{\bibfnamefont{I.~B.} \bibnamefont{Spielman}},
  \bibinfo{author}{\bibfnamefont{W.~D.} \bibnamefont{Phillips}},
  \bibnamefont{and} \bibinfo{author}{\bibfnamefont{J.~V.} \bibnamefont{Porto}},
  \bibinfo{journal}{Phys. Rev. Lett.} \textbf{\bibinfo{volume}{98}},
  \bibinfo{eid}{080404} (\bibinfo{year}{2007}).

\bibitem[{\citenamefont{Elstner and Monien}(1999)}]{elstner:59:12184}
\bibinfo{author}{\bibfnamefont{N.}~\bibnamefont{Elstner}} \bibnamefont{and}
  \bibinfo{author}{\bibfnamefont{H.}~\bibnamefont{Monien}},
  \bibinfo{journal}{Phys. Rev. B} \textbf{\bibinfo{volume}{59}},
  \bibinfo{pages}{12184} (\bibinfo{year}{1999}).

\bibitem[{Not({\natexlab{f}})}]{Note6}
\bibinfo{note}{Additional information can be obtained from the pair correlation
  function $G_2(x,x')$ in deep lattices with one particle fixed in the middle
  of a specific site. As seen before, the correlation vanishes at that site for
  filling factor $\nu=1$. For noninteger filling ($6 \leq N \leq 9$) the height
  of the correlation function varies for each site. Thus, regarding the integer
  number of particles, at least some particles must be delocalized between the
  sites.}

\bibitem[{Not({\natexlab{g}})}]{Note7}
\bibinfo{note}{The first and second excited state, the third and fourth excited
  state as well as the sixth and seventh excited state are exactly degenerate.}

\bibitem[{\citenamefont{Batrouni et~al.}(2002)\citenamefont{Batrouni, Rousseau,
  Scalettar, Rigol, Muramatsu, Denteneer, and Troyer}}]{Batrouni}
\bibinfo{author}{\bibfnamefont{G.~G.} \bibnamefont{Batrouni}},
  \bibinfo{author}{\bibfnamefont{V.}~\bibnamefont{Rousseau}},
  \bibinfo{author}{\bibfnamefont{R.~T.} \bibnamefont{Scalettar}},
  \bibinfo{author}{\bibfnamefont{M.}~\bibnamefont{Rigol}},
  \bibinfo{author}{\bibfnamefont{A.}~\bibnamefont{Muramatsu}},
  \bibinfo{author}{\bibfnamefont{P.~J.~H.} \bibnamefont{Denteneer}},
  \bibnamefont{and} \bibinfo{author}{\bibfnamefont{M.}~\bibnamefont{Troyer}},
  \bibinfo{journal}{Phys. Rev. Lett.} \textbf{\bibinfo{volume}{89}},
  \bibinfo{pages}{117203} (\bibinfo{year}{2002}).

\end{thebibliography}
\end{document}